\documentclass[aps,twocolumn,tightenlines,preprintnumbers,nofootinbib,superscriptaddress,byrevtex,floatfix,showpacs]{revtex4}
\usepackage{amsmath,amssymb}
\usepackage{wasysym}
\usepackage{graphicx}
\usepackage{color}
\usepackage{comment}
\usepackage{multirow}
\begin{document}

% Greek Letters
\def\a{{\alpha}}
\def\b{{\beta}}
\def\d{{\delta}}
\def\D{{\Delta}}
\def\e{{\epsilon}}
\def\g{{\gamma}}
\def\G{{\Gamma}}
\def\k{{\kappa}}
\def\l{{\lambda}}
\def\L{{\Lambda}}
\def\m{{\mu}}
\def\n{{\nu}}
\def\w{{\omega}}
\def\O{{\Omega}}
\def\S{{\Sigma}}
\def\s{{\sigma}}
\def\t{{\tau}}
\def\th{{\theta}}
\def\x{{\xi}}

\def\ol#1{{\overline{#1}}}
\def\slash#1{{#1\hskip-0.4em /}}

%slash's
\def\Dslash{D\hskip-0.65em /}
\def\dslash{{\partial\hskip-0.5em /}}
\def\vslash{{\rlap \slash v}}
\def\qbar{{\overline q}}
\def\pslash{p\hskip-0.4em /}
\def\ppslash{p\hskip-0.4em /^\prime}

% Jargon
\def\CPT{{$\chi$PT}}
\def\QCPT{{Q$\chi$PT}}
\def\PQCPT{{PQ$\chi$PT}}
\def\tr{\text{tr}}
\def\str{\text{str}}
\def\diag{\text{diag}}
\def\order{{\mathcal O}}
\def\vit{{\it v}}
\def\vD{\vit\cdot D}
\def\am{\alpha_M}
\def\bm{\beta_M}
\def\gm{\gamma_M}
\def\smb{\sigma_M}
\def\smt{\overline{\sigma}_M}
\def\tb{{\tilde b}}

\def\mc#1{{\mathcal #1}}

% Fields
\def\Bbar{\overline{B}}
\def\Tbar{\overline{T}}
\def\cBbar{\overline{\cal B}}
\def\cTbar{\overline{\cal T}}
\def\pq{(PQ)}

\def\eqref#1{{(\ref{#1})}}

\unitlength=1mm
\def\pq{{p\cdot q}}
\def\tel{{\tau_{el}}}
\def\Tmn{{T_{\mu\nu}}}
\def\DOnemn{{\left( g_{\mu\nu} - \frac{q_\mu q_\nu}{q^2} \right)}}
\def\DTwomn{{\frac{1}{M^2} \left( p_\mu - \frac{\pq}{q^2} q_\mu \right) 
	\left(p_\nu - \frac{\pq}{q^2} q_\nu \right) }}
\def\dTwomn{{\frac{1}{M^2} \left( p_\mu p_\nu - \frac{\pq}{q^2}(p_\mu q_\nu + p_\nu q_\mu) +\frac{(\pq)^2}{q^2} g_{\mu\nu} \right)}}
\def\Tr{\mathbf{Tr}}
\def\dMnpg{\delta M_{p-n}^\gamma}

{\count255=\time\divide\count255 by 60 \xdef\hourmin{\number\count255}
  \multiply\count255 by-60\advance\count255 by\time
  \xdef\hourmin{\hourmin:\ifnum\count255<10 0\fi\the\count255}}

\title{The Electromagnetic Self-Energy Contribution to $M_p - M_n$\\
and the Isovector Nucleon Magnetic Polarizability}

\author{Andr\'{e} Walker-Loud} 
%	\email[]{awalker-loud@lbl.gov}
	\affiliation{Lawrence Berkeley National Laboratory, Berkeley, CA 94720, USA}
	\affiliation{Department of Physics, University of California, Berkeley, CA 94720, USA}
\author{Carl E. Carlson} 
%	\email[]{carlson@physics.wm.edu}
	\affiliation{Department of Physics, College of William and Mary, Williamsburg, VA 23187-8795.}
\author{Gerald A. Miller} 
%	\email[]{miller@phys.washington.edu}
	\affiliation{Lawrence Berkeley National Laboratory, Berkeley, CA 94720, USA}
	\affiliation{Department of Physics, University of Washington, Seattle, WA 98195-1560.}

\begin{abstract}
We update the determination of the isovector nucleon electromagnetic self-energy, valid to leading order in QED.
A technical oversight in the literature concerning the elastic contribution to Cottingham's formula is corrected and modern knowledge of the structure functions is used to precisely determine the inelastic contribution.
We find $\delta M_{p-n}^\gamma = 1.30(03)(47)$~MeV.
The largest uncertainty arises from a subtraction term required in the dispersive analysis, which can be related to the isovector magnetic polarizability.
With plausible model assumptions, we can combine our calculation with additional input from lattice QCD to constrain this polarizability as: $\beta_{p-n} = -0.87(85)\times 10^{-4} \textrm{fm}^{3}$.
\end{abstract}

\date{\today\quad\hourmin}

\maketitle

%%%%%%%%%%%%%%%%%%%%%%%%%%%%%%%%%%%%%%%%%%%%%%%%
%\textbf{\textit{Introduction}}-- 
%%%%%%%%%%%%%%%%%%%%%%%%%%%%%%%%%%%%%%%%%%%%%%%%
Given only electrostatic forces, one would predict that the proton is more massive than the neutron but the opposite actually occurs~\cite{Mohr:2008fa,Nakamura:2010zzi,Codata:2012}:
\begin{equation}
\label{eq:dMexp}
	M_n - M_p = 1.29333217(42) \textrm{ MeV}\, .
\end{equation}
Before we knew of quarks and gluons there were many attempts to explain this contradiction, see Ref.~\cite{Zee:1971df} for a review.
We now know there are two sources of isospin breaking in the standard model, the masses of the $up$ and $down$ quarks
as well as the electromagnetic interactions between quarks governed by the charge operator. 
The effects of the mass difference between down and up quarks are  larger and of the opposite sign than those of electromagnetic effects, see the reviews~\cite{Gasser:1982ap,Miller:1990iz,Miller:2006tv}. 
The net result of the quark mass difference and electromagnetic effects  is well known, Eq.~\eqref{eq:dMexp}, but our ability to disentangle the contributions from these two sources remains poorly constrained.

In contrast, lattice QCD calculations have matured significantly.  There are now calculations performed with the light quark masses at or near their physical values~\cite{Aoki:2008sm,Bazavov:2009bb,Aoki:2009ix,Durr:2010vn,Durr:2010aw}, reproducing the ground state hadron spectrum within a few percent.
These advances have allowed for calculations to begin including explicit isospin breaking effects from both the quark masses~\cite{Beane:2006fk,WalkerLoud:2009nf,Blum:2010ym,WalkerLoud:2010qq,deDivitiis:2011eh} and electromagnetism~\cite{Blum:2010ym,Torok:2010zz,Portelli:2010yn,Tiburzi:2011vk,Ukita:2011vk}.
While the lattice calculations of $m_d - m_u$ effects are robust, the contributions from electromagnetism are less mature and suffer from larger systematics, due in large part to the disparity between the photon mass and a typical hadronic scale.
Improved knowledge of $m_d - m_u$ and its effects in nucleons will enhance the ability to use effective field theory to compute a variety of isospin-violating (charge asymmetric) effects in nuclear reactions~\cite{Miller:2006tv,vanKolck:1996rm,vanKolck:2000ip,Stephenson:2003dv,Opper:2003sb,Gardestig:2004hs,Nogga:2006cp}.

An application~\cite{Gasser:1974wd} of the Cottingham sum rule~\cite{Cottingham:1963zz}, which relates the electromagnetic self-energy of the nucleon to measured elastic and inelastic cross sections, gives the result 
$\dMnpg =0.76\pm 0.30$ MeV. 
Given the high  present interest in the precise value of $\dMnpg$ and its many possible implications, it is worthwhile to revisit this result. Many high quality electron scattering experiments have been performed since 1975 and there have also been theoretical advances.
The central aim of this work is to provide a modern, robust evaluation of $\dMnpg$.
We will show the precision of this effort is severely limited by our knowledge of the required subtraction function.
Given plausible model assumptions, this limitation is translated into our knowledge of the isovector nucleon magnetic polarizability, $\b_{p-n} = \b_M^p - \b_M^n$, for which even the sign is presently unknown~\cite{Griesshammer:2012we}.

%%%%%%%%%%%%%%%%%%%%%%%%%%%%%%%%%%%%%%%%%%%%%%%%
\textit{Cottingham's sum rule--}
%%%%%%%%%%%%%%%%%%%%%%%%%%%%%%%%%%%%%%%%%%%%%%%%
In perturbation theory, the electromagnetic self-energy of the nucleon,
$\d M^\g$, can be related to the spin averaged forward Compton scattering tensor
\begin{equation}\label{eq:Tmunu}
T_{\mu\nu} = \frac{i}{2} \sum_\s \int d^4 \xi\ e^{i q\cdot \xi}
	\langle p \s | T \left\{ J_\mu(\xi) J_\nu(0) \right\} | p \s \rangle\, ,
\end{equation}
integrated with the photon propagator over space-time
\begin{align}\label{eq:dM}
\d M^\g &= \frac{i}{2M} \frac{\a}{(2\pi)^3}
	\int_R d^4 q \frac{T_\mu^\mu(p,q)}{q^2 + i\e}\, ,
\end{align}
where we work in the nucleon rest frame $p^\mu = (M,\mathbf{0})$,
$\a = e^2 / 4\pi$ 
and the subscript $R$ implies the integral has been renormalized.
Performing a Wick rotation of the integration contour to imaginary photon energy, the nucleon self-energy can be related to the structure functions arising from the scattering of space-like photons through dispersion theory, giving rise to what is known as Cottingham's formula (the Cottingham sum rule)~\cite{Cini:1959,Cottingham:1963zz}.
In principle, this allows the integral in Eq.~\eqref{eq:dM} to be computed in a model independent fashion with input from experimental data.
There are a few issues  which complicate the realization of this method: a subtracted dispersive analysis is required introducing an unknown subtraction function~\cite{Harari:1966mu,Abarbanel:1967zza};
the integral in Eq.~\eqref{eq:dM} diverges logarithmically in the ultra-violet  region and requires renormalization~\cite{Collins:1978hi}.
We review these issues briefly.

Lorentz invariance significantly constrains the form of $T_{\mu\nu}$, for which there are two common parameterizations,
\begin{subequations}
\begin{align}
\label{eq:Tmunu_a}
T_{\mu\nu}(p,q) &= 
	-D^{(1)}_{\mu\nu} T_1 (\nu,-q^2 ) 
	+D^{(2)}_{\mu\nu} T_2 (\nu,-q^2 ) 
\\
\label{eq:Tmunu_b}
	&= \phantom{+}d^{(1)}_{\mu\nu}\, 
		q^2 t_1(\nu,-q^2 ) 
		-d^{(2)}_{\mu\nu} q^2 t_2(\nu,-q^2 ) 
\end{align}
\end{subequations}
where $\pq = M\nu$ and 
\begin{align}
	d^{(1)}_{\mu\nu}&=  D^{(1)}_{\mu\nu} = g_{\mu\nu} - \frac{q_\mu q_\nu}{q^2}\, ,
\nonumber\\
	d^{(2)}_{\mu\nu} &= \dTwomn\, ,
\nonumber\\
	D^{(2)}_{\mu\nu} &= \DTwomn\, .
\end{align}
Performing the Wick rotation $\nu \rightarrow i\nu$ and the variable transformation $Q^2 = \mathbf{q}^2 +\nu^2$, the self-energy becomes
\begin{align}
\d M^\g =& \frac{\a}{8\pi^2} \int_{0}^{\L^2} {\hskip-0.8em} dQ^2 \int_{-Q}^{+Q} {\hskip-0.8em}d\nu 
	\frac{\sqrt{Q^2 - \nu^2}}{Q^2} 
	\frac{T_\mu^\mu}{M}
	+\d M^{ct}(\L)
\end{align}
where $\d M^{ct}(\L)$ derives from counterterms required for renormalization~\cite{Collins:1978hi} and the Lorentz contracted Compton tensor is
\begin{subequations}
\label{eq:dMPreDispersive}
\begin{align}
T_\mu^\mu &= 
\label{eq:dMPreDispersivea}
	-3\, T_1(i\nu,Q^2) + \left( 1 - \frac{\nu^2}{Q^2} \right) T_2 (i\nu,Q^2)\, ,
\\
\label{eq:dMPreDispersiveb}
	&= -3Q^2\, t_1(i\nu,Q^2) + \left( 1 +2\frac{\nu^2}{Q^2} \right) Q^2 t_2(i\nu,Q^2)\, .
\end{align}
\end{subequations}
The scalar functions $(T_i,t_i)$ can be evaluated using a dispersive analysis.
It is known the $(T_1,t_1)$ functions require a subtracted dispersive analysis while the $(T_2,t_2)$ functions can be evaluated with an unsubtracted dispersion relation~\cite{Harari:1966mu}.
In Ref.~\cite{Gasser:1974wd}, it was claimed the elastic contributions to $t_1$ could be evaluated with an unsubtracted dispersive analysis.
However, performing an unsubtracted dispersive analysis of the elastic contributions to Eqs.~\eqref{eq:dMPreDispersive} 
by  inserting a complete set of elastic  states into Eq.~\eqref{eq:Tmunu}, leads to  inconsistent results: 
\begin{widetext}
\begin{subequations}
\begin{align}
\label{eq:dMelastica}
\d M^{el}_{unsub,a} &= 
	\frac{\a}{\pi} \int_{0}^{\L^2} dQ
		\bigg\{
			\left[ G_E^2(Q^2) -2\tel G_M^2(Q^2) \right] \frac{(1+\tel)^{3/2} - \tel^{3/2} - \frac{3}{2}\sqrt{\tel}}{1+\tel}
			-\frac{3}{2} G_M^2(Q^2) \frac{\tel^{3/2}}{1+\tel}
		\bigg\}\, ,
\\
\label{eq:dMelasticb}
\d M^{el}_{unsub,b} &= 
	\frac{\a}{\pi} \int_{0}^{\L^2} dQ
		\bigg\{ 
			\left[ G_E^2(Q^2) -2\tel\, G_M^2(Q^2) \right] \frac{(1+\tel)^{3/2} - \tel^{3/2}}{1+\tel}
			+ 3G_M^2(Q^2) \frac{\tel^{3/2}}{1+\tel}
		\bigg\}\, ,
\end{align}
\end{subequations}
\end{widetext}
%%gm1-17
with $\tel \equiv \frac{Q^2}{4M^2}$.
If both parameterizations of the elastic contribution were to satisfy unsubtracted dispersion relations, the following positive-definite integral would have to vanish
\begin{align}
\frac{3\a}{2\pi} \int_0^\infty {\hskip-0.6em}dQ \sqrt{\tel} \frac{G_E^2(Q^2) +\tel G_M^2(Q^2)}{1+\tel}\, .
\end{align}
Equating Eqs.~\eqref{eq:Tmunu_a} and \eqref{eq:Tmunu_b} allows one to solve for $T_i$ in terms of $t_i$ and vice versa, and to demonstrate that if the elastic contributions to $T_1 (t_1)$ do not satisfy an unsubtracted dispersive analysis, then neither will the elastic contributions to $t_1 (T_1)$.
Eq.~\eqref{eq:dMelasticb} was used in Ref.~\cite{Gasser:1974wd} and is often quoted as the elastic contribution to the nucleon self-energy.

Starting from either Eqs.~\eqref{eq:dMPreDispersivea} or \eqref{eq:dMPreDispersiveb}, performing a subtracted dispersive analysis of $(T_1,t_1)$, and an unsubtracted analysis of $(T_2,t_2)$, using a mass-independent renormalization scheme (dimensional regularization) one arrives at~\cite{Collins:1978hi}
\begin{align}\label{eq:dMg}
\d M^\g = \d M^{el} + \d M^{inel} + \d M^{sub} + \d \tilde{M}^{ct}\, ,
\end{align}
with
% ELASTIC
\begin{multline}
\label{eq:dMel}
\d M^{el} = \frac{\a}{\pi} \int_{0}^{\L_0^2} 
	{\hskip-0.6em}dQ
	\bigg\{ 
	\frac{3\sqrt{\tel}G_M^2}{2(1+\tel)}\, 
	+
	\frac{\left[ G_E^2 -2\tel\, G_M^2 \right]}{1+\tel}
\\%\nonumber\\&
	\times
	\bigg[(1+\tel)^{3/2} -\tel^{3/2} -\frac{3}{2}\sqrt{\tel} \bigg]
	\bigg\}\, ,
\end{multline}
% INELASTIC
\begin{align}
\label{eq:dMinel}
\d M^{inel} &= \frac{\a}{\pi} 
		\int_0^{\L_0^2} \frac{dQ^2}{2Q} \int_{\nu_{th}}^\infty
		d\nu 
		\bigg\{
\nonumber\\&
		\phantom{+}
		\frac{3F_1(\nu,Q^2)}{M} \bigg[
			\frac{\t^{3/2} -\t\sqrt{1+\t} +\sqrt{\tau}/2 }{\t}
		\bigg]
\nonumber\\&
		+\frac{F_2(\nu,Q^2)}{\nu} \bigg[ 
			(1+\t)^{3/2} -\t^{3/2} -\frac{3}{2}\sqrt{\t}
		\bigg] \bigg\}\, ,
\end{align}
where $\t = \nu^2 / Q^2$, $F_i(\nu,Q^2)$ are the standard nucleon structure functions and $\nu_{th} = m_\pi + (m_\pi^2 +Q^2)/2M$;
% SUB
\begin{align}
\label{eq:dMsub}
\d M^{sub} &=	-\frac{3\a}{16\pi M}\int_0^{\L_0^2} dQ^2\ T_1(0,Q^2)\, ,
\end{align}
and
% CT
\begin{align}
\d \tilde{M}^{ct} &= -\frac{3\a}{16\pi M} \int_{\L_0^2}^{\L_1^2} dQ^2
	\sum_i C_{1,i} \langle \mc{O}^{i,0} \rangle\, ,
\end{align}
where $C_{1,i}$ are Wilson coefficients determined from the operator product expansion of the counterterms~\cite{Collins:1978hi}.
The UV divergence has been entirely cancelled by the counterterm and $\d \tilde{M}^{ct}$ is a remaining finite contribution with residual scale dependence.  The scales $\L_0$ and $\L_1$ can be chosen arbitrarily provided their values are in the asymptotic scaling region.
Restricting our attention to the isospin breaking contribution, with $2\d = m_d - m_u$
%\begin{multline}
%\delta \tilde M^{ct}_{p-n} = 
%	-\frac{3\alpha}{8\pi M }	\ln\left(  \frac{\Lambda_1^2}{\Lambda_0^2}  \right)
%	\frac{e_u^2 m_u - e_d^2  m_d }{\delta}  
%\\
%	\times
%	\langle p | \delta ( \bar u u - \bar d d)  | p \rangle
%\end{multline}
\begin{multline}
\delta \tilde M^{ct}_{p-n} = 
	3\alpha	\ln\left(  \frac{\Lambda_0^2}{\Lambda_1^2}  \right)
	\frac{e_u^2 m_u - e_d^2  m_d }{8\pi M\delta}  
%\\
%	\times
	\langle p| \delta(\bar u u - \bar d d) | p\rangle
\end{multline}
%\begin{align}
%	\d \tilde{M}_{p-n}^{ct} = \frac{\a}{16\pi M}
%		\ln \left( \frac{\L_1^2}{\L_0^2} \right)
%		\frac{m_u -\frac{2}{3}\d}{\d}
%		\langle \d(\bar{u}u - \bar{d}d) \rangle\, .
%\end{align}
with $e_u = 2/3$ and $e_d = -1/3$.
In QCD, $m_{u,d} \sim \d$, so the entire contribution is numerically second order in isospin breaking, $\mc{O}(\a \d)$, and for practical purposes can be neglected~\cite{Collins:1978hi}.
Estimating the size of this term, with $\L_1^2 = 100 \textrm{ GeV}^2$, $\L_0^2 = 2 \textrm{ GeV}^2$ yields $| \d \tilde{M}_{p-n}^{ct} | < 0.02$~MeV.

The remaining contribution to the self-energy is the subtraction term, which can not be directly related to experimentally measured cross sections.
We now have a better theoretical understanding of this term enabling a more robust determination of its contribution than has been previously made. 
While the function is not known, the low and high $Q^2$ limits can be determined in a model independent fashion; the asymptotic region is constrained by the operator product expansion (OPE) to scale as $\lim_{Q \rightarrow \infty}T_1(0,Q^2) \sim 1/Q^2$~\cite{Collins:1978hi} while the low $Q^2$ limit is fixed by non-relativistic QED~\cite{Pachucki:1996zza,Pineda:2002as,Pineda:2004mx,Hill:2011wy,Carlson:2011zd}
\begin{multline}
\label{eq:T1sub}
T_1(0,Q^2) = 2\k (2+\k)
-Q^2 \bigg\{
		\frac{2}{3} \left[ (1+\k)^2 r_M^2 - r_E^2 \right]
\\
		+\frac{\k}{M^2}
%inelastic
	-2 M  \frac{\b_M}{\a}
	\bigg\}
	+\mc{O}(Q^4)
\, ,
\end{multline}
where $\k \equiv F_2(0)$ is the anomalous magnetic moment, $r_{E} (r_{M})$ is proportional to the slope of the electric (magnetic) form factor and commonly denoted as the nucleon electric (magnetic) charge radius and $\b_M$ is the magnetic polarizability.

A direct evaluation of Eq.~\eqref{eq:dMsub} with Eq.~\eqref{eq:T1sub} diverges quadratically resulting in an uncontrolled uncertainty.
However, the displayed $Q^2$ dependence is not of the form required by the OPE, so we are necessarily led to introduce model dependence.
The first few terms in Eq.~\eqref{eq:T1sub} are recognized as the low-$Q^2$ expansion of elastic form factors and the magnetic polarizability term is the leading inelastic contribution.
In evaluating the elastic contributions to $T_{\mu\nu}$ only the elastic u-spinors need be used in the dispersion relation. 
If one uses the full Feynman propagator in the full amplitude, a procedure known to be correct in the point-limit (as for the electron), 
and vertex functions with ordinary $F_1$ and $F_2$ form factor contributions, 
then  the specific  elastic terms of Eq.~\eqref{eq:T1sub} would arise~\cite{Carlson:2011zd,Carlson:2011dz}.
This suggests a re-summation in which one uses the appropriate elastic form factors.
The inelastic contribution can be multiplied by a dipole form factor $(1+Q^2 / m_0^2)^{-2}$, such that it has the correct asymptotic limits as $Q^2 \rightarrow 0,\infty$.
The parameter $m_0^2$ should be a typical hadronic scale and we will take $m_0^2 = 0.71$ GeV$^2$.
The subtraction term is then approximated by two pieces which have the correct low and high $Q^2$ limiting behavior,
\begin{align}
\label{eq:T1sub_model}
T_1(0,Q^2) \simeq&\ 2G_M^2(Q^2) 
	- 2 F_1^2(Q^2)
\nonumber\\&
	+Q^2 2 M  \frac{\b_M}{\a} \left( \frac{m_0^2}{m_0^2 + Q^2} \right)^2\, ,
\end{align}
leading to the convenient separation
\begin{subequations}
\label{eq:dMsub2}
\begin{align}
\label{eq:dMsub_el}
\d M^{sub}_{el} &= -\frac{3\a}{16\pi M} \int_0^{\L_0^2} dQ^2 \bigg[
	2 G_M^2 -2 F_1^2
	\bigg]\, ,
\\
\label{eq:dMsub_inel}
\d M^{sub}_{inel} &= -\frac{3\b_M}{8\pi } \int_0^{\L_0^2} dQ^2
	Q^2 \left( \frac{m_0^2}{m_0^2 + Q^2} \right)^2\, .
\end{align}
\end{subequations}
The second term, generated using the model assumptions described above, will cause the largest uncertainties, as we show below.

%%%%%%%%%%%%%%%%%%%%%%%%%%%%%%%%%%%%%%%%%%%%%%%%
\textit{Evaluation of contributions--}
%%%%%%%%%%%%%%%%%%%%%%%%%%%%%%%%%%%%%%%%%%%%%%%%
In all subsequent evaluations, we take $\L_0^2 = 2 \textrm{ GeV}^2$ for our central values and the range $1.5^2 < \L_0^2 < 2.5 \textrm{ GeV}^2$ to estimate uncertainties.
We begin with an evaluation of the elastic contribution, Eq.~\eqref{eq:dMel}.
The form factors are well measured over the kinematic range required by the integrals, which are represented by a number of analytic fits.  The elastic contributions converge well at the upper limit, which may be taken to infinity with negligible error.
Using the Kelly parameterization of the form factors~\cite{Kelly:2004hm}, or an updated version~\cite{Bradford:2006yz,Arrington:2006hm,Venkat:2010by}, the elastic contribution is given by
\begin{equation}
	\label{eq:dMel_num}
	\left.
	\d M^{el} \right|_{p-n} = 1.39(02) \textrm{ MeV}\, .
\end{equation}
The uncertainty is determined through an uncorrelated Monte-Carlo evaluation of the fit parameters in the parametrization.%
%FOOTNOTE
%\footnote{We thank J.~Arrington for help with this estimate.} 
%
 It is also interesting to note, that if the simple dipole parameterization of the form factors is used, the same value within the quoted uncertainty is obtained.

In the inelastic contribution, Eq.~\eqref{eq:dMinel}, most of the support for the integrals lies in the resonance region, where there are good data from JLab, and there are analytic fits valid in the resonance region for both the neutron and proton structure functions from Bosted and Christy~\cite{Bosted:2007xd,Christy:2007ve} (we also remind the reader the neutron functions are determined from deuterium-Compton scattering with the additional uncertainties captured in the coefficients of the neutron functions, and propagated into our uncertainties through a Monte-Carlo treatment).  Their quoted range of validity includes $Q^2$ up to $8$ GeV$^2$ and $W$ up to $3.1$ GeV $(W^2 = M^2 +2M\nu - Q^2$).
To extend the $W$ range, we use the parameterizations of Refs.~\cite{Capella:1994cr,Sibirtsev:2010zg} which fit proton structure functions in the diffraction region using forms recognizable as Pomeron and rho meson Regge trajectories.  The former is isoscalar and the latter isovector, so we have a straightforward extension to the neutron case.
Taking $\L_0^2 = 2 \textrm{ GeV}^2$ and $W_{trans} = 3.1$~GeV as the transition between the two parameterizations, the inelastic contribution is%
%FOOTNOTE
%\footnote{We thank P.~Bosted and E.~Christy for access to their analysis code enabling a determination of the uncertanties.} 
%
\begin{equation}
\label{eq:dMinel_num}
	\left. 
	\d M^{inel} \right|_{p-n} = 0.057(16) {\rm \ MeV}	.
\end{equation}
The uncertainties are estimated by the range of $\L_0^2$ given above as well as varying the transition value of $W$ between $2.5 < W_{trans} < 3.5$~GeV.  These two variations dominate the uncertainty estimate.
The numerical integration is insensitive to the upper limit of the $W$ integration through $W_{max} \sim 200$~GeV (or $x \sim 10^{-4}$).

We are left with the subtraction terms.  Using the model assumptions described above, the contribution from the elastic subtraction term, Eq.~\eqref{eq:dMsub_el}, is 
\begin{equation}
\label{eq:dMsub_el_num}
	\left.
	\d M^{sub}_{el} \right|_{p-n} = -0.62(02) \textrm{ MeV}
\end{equation}
It is interesting to note the sum of Eqs.~\eqref{eq:dMel_num} and \eqref{eq:dMsub_el_num} is surprisingly close to that of Ref.~\cite{Gasser:1974wd} (although the individual proton and neutron elastic self energies are different).

The most troublesome contribution to evaluate is that of the inelastic subtraction term, Eq.~\eqref{eq:dMsub_inel}.
This contribution is proportional to the isovector nucleon magnetic polarizability $\b_{p-n}$.
%Presently, even the sign of $\b_{p-n}$ is unknown.  
The determination of this isovector quantity was part of the motivation for the recent deuterium Compton scattering experiment, MAX-Lab at Lund~\cite{Lundin:2002jy}, for which we are still awaiting results.
The HIGS experiment~\cite{Weller:2009zza} at TUNL will also help determine this quantity.
From chiral perturbation theory, one expects the isovector polarizabilities to be small; the leading contribution to the polarizabilities occurs at order $P^3$ and these are purely isoscalar.  
The isovector contributions  arise at order $P^4$ and are suppressed in the chiral power counting~\cite{Beane:2002wn}.
A recent review provides the conservative estimate $\b_{p-n} = -1\pm 1 \times 10^{-4} \textrm{ fm}^3$~\cite{Griesshammer:2012we}.
Using this in Eq.~\eqref{eq:dMsub_inel} provides the determination
\begin{equation}
	\label{eq:dMsub_inel_num}
	\left.
	\d M^{sub}_{inel} \right|_{p-n} = 0.47 \pm 0.47 \textrm{ MeV}\, ,
\end{equation}
(a smaller value of $m_0^2$ would reduce these values).

Adding all the various contributions, Eqs.~\eqref{eq:dMel_num}, \eqref{eq:dMinel_num}, \eqref{eq:dMsub_el_num} and \eqref{eq:dMsub_inel_num}, we arrive at
\begin{equation}
	\left.
	\d M^\g \right|_{p-n} = 1.30(03)(47) \textrm{ MeV}\, ,
\end{equation}
where the second uncertainty arises from the inelastic contribution to the subtraction term.
Clearly, any improvement in our knowledge of $\b_{p-n}$ will significantly improve our ability to determine the electromagnetic contribution to $M_p - M_n$.

%%%%%%%%%%%%%%%%%%%%%%%%%%%%%%%%%%%%%%%%%%%%%%%%
\textit{The isovector magnetic polarizability--}
%%%%%%%%%%%%%%%%%%%%%%%%%%%%%%%%%%%%%%%%%%%%%%%%
Within the model assumptions used to arrive at Eqs.~\eqref{eq:dMsub2}, we can combine the experimental value for $M_n - M_p$ with lattice QCD determinations of the $m_d - m_u$ contribution.
There are three published numbers from lattice QCD~\cite{Beane:2006fk,Blum:2010ym,deDivitiis:2011eh}, which are uncorrelated.  For each result, we combine the quoted uncertainties in quadrature, and then perform a simple weighted mean, arriving at
\begin{equation}
\label{eq:dM_latt}
	\left.
	\d M^{latt}_{m_d -m_u}\right|_{p-n} = -2.53(40) \textrm{ MeV}\, .
\end{equation}
Combining this with Eqs.~\eqref{eq:dMexp}, \eqref{eq:dMsub_inel}, \eqref{eq:dMel_num}, \eqref{eq:dMinel_num}, \eqref{eq:dMsub_el_num} and our value for $m_0^2$, we find
\begin{equation}
\b_{p-n} = -0.87(85) \times 10^{-4} \textrm{ fm}^3\, ,
\end{equation}
in good agreement with current estimates~\cite{Griesshammer:2012we}.

%%%%%%%%%%%%%%%%%%%%%%%%%%%%%%%%%%%%%%%%%%%%%%%%
\textit{Model independence--} 
%%%%%%%%%%%%%%%%%%%%%%%%%%%%%%%%%%%%%%%%%%%%%%%%
One can infer the nucleon isovector electromagnetic self-energy without recourse to models by utilizing the known mass splitting, Eq.~\eqref{eq:dMexp}, combined with the lattice QCD determination of the of the contribution from $m_d - m_u$, Eq.~\eqref{eq:dM_latt},
\begin{equation}
\d M_{p-n}^\g = 1.24(40) \textrm{ MeV}\, .
\end{equation}
Combined with Eqs.~\eqref{eq:dMel_num} and \eqref{eq:dMinel_num}, this can be translated into a model-independent bound on the unknown subtraction function
\begin{equation}
\frac{3\a}{16\pi M} \int_0^{\L_0^2} 
	{\hskip-0.6em}
	dQ^2\ 
	T_1^{p-n}(0,Q^2)
	= 0.21(02)(40) \textrm{ MeV}\, .
\end{equation}
This is compared with Eqs.~\eqref{eq:dMsub_el_num} and \eqref{eq:dMsub_inel_num} which give $0.15(02)(47)$~MeV for the same quantity.
This bound demonstrates that our treatment of the subtraction function, while not model-independent, is also not wildly speculative, but in agreement with the combined constraint of experiment and lattice QCD.

%%%%%%%%%%%%%%%%%%%%%%%%%%%%%%%%%%%%%%%%%%%%%%%%
\bigskip
\textit{Conclusions}-- 
%%%%%%%%%%%%%%%%%%%%%%%%%%%%%%%%%%%%%%%%%%%%%%%%
We have provided a modern and robust determination of the isovector electromagnetic self-energy contribution, $\d M^\g_{p-n} = 1.30(03)(47)$.
A technical oversight in the evaluation of the elastic contribution was highlighted resulting in a larger central value than previously obtained~\cite{Gasser:1974wd}.
%
%{\color{blue}
Modern knowledge of the structure functions was used to constrain the elastic and inelastic contributions, reducing the uncertainty from these sources by an order of magnitude ($\pm 0.30$~MeV~\cite{Gasser:1974wd} compared to our $\pm0.03$~MeV).
However, %} 
a careful analysis of the subtraction function has yielded an overall larger uncertainty than previously recognized.
The larger central value suggests a larger contribution to $M_p - M_n$ from $m_d - m_u$, consistent with expectations from lattice QCD, thus impacting the phenomenology of Refs.~\cite{vanKolck:1996rm,vanKolck:2000ip,Stephenson:2003dv,Opper:2003sb,Gardestig:2004hs,Nogga:2006cp}.
With plausible model assumptions and additional input from lattice QCD, this knowledge can be used to provide a competitive estimate of the nucleon isovector magnetic polarizability, albeit still with a 100\% uncertainty.
Alternatively, a bound can be placed on the unknown subtraction function, which can not otherwise be determined, and lends further support for our determination of $\b_{p-n}$.

\section*{Acknowledgements}
We would like to thank many for helpful conversations and correspondence: J.~Arrington for help with the Monte Carlo evaluation of the fit parameters, P.~Bosted, E.~Christy for access to their analysis code enabling a determination of the uncertainties to the inelastic contribution in Eq.~\eqref{eq:dMinel_num}, W.~Detmold, J.~Gasser, M.~Golterman, B.~Holstein, H.~Leutwyler, S.~Paris, D.~Phillips, S.~Syritsyn, B.~Tiburzi, M.~Vanderhaeghen, L.~Yaffe and F.~Yuan.
GAM would like to thank the hospitality of LBNL where some of this work was carried out.
This work was supported in part by the National Science Foundation under Grant PHY-0855618
and by the U.S. Department of Energy under Grant Nos. DE-FG02-97ER-41014 and DE-AC02-05CH11231.

\bibliography{em_selfEnergy}

\end{document}